\documentclass[lettersize,journal]{IEEEtran}
\usepackage{amsmath,amsfonts}
\usepackage{amssymb, amsthm}
\usepackage{algorithmic}
\usepackage{algorithm}
\usepackage{array}
\usepackage[colorlinks=false,linkbordercolor={1 0 0},citebordercolor={0 1 0},urlbordercolor={1 0 0}]{hyperref}
\usepackage[caption=false,font=normalsize,labelfont=sf,textfont=sf]{subfig}
\usepackage{textcomp}
\usepackage{stfloats}
\usepackage{url}
\usepackage{verbatim}
\usepackage{graphicx}
\usepackage{cite}
\newtheorem{lemma}{Lemma}
\newtheorem{proposition}{Proposition}
\newtheorem{theorem}{Theorem}
\newtheorem{remark}{Remark}
\newtheorem{corollary}{Corollary}
\usepackage{comment}

\begin{document}

\title{Optimal Sampling and Transmission Policy for Information Freshness in Gossip Networks}

\author{Irtiza~Hasan,~\IEEEmembership{Graduate Student Member,~IEEE,}~and~Ahmed~Arafa,~\IEEEmembership{Member,~IEEE}
\thanks{The authors are with the Department of Electrical and Computer Engineering, University of North Carolina at Charlotte, Charlotte, NC 28223 USA; e-mails: ihasan@charlotte.edu; aarafa@charlotte.edu.}
\thanks{This work was supported by the U.S. National Science Foundation under Grant ECCS 21-46099.}}



\maketitle

\begin{abstract}
We study the optimal sampling and transmission policies for a transmitter communicating with $n$ \emph{gossiping} receivers with the objective of tracking a source's status over time using a semantic freshness metric: Version Age of Information (VAoI). The transmitter observes the version changes in the source, and can acquire a new information version by paying a sampling cost. The receivers \emph{gossip}, allowing information exchange among each other without direct transmission. Gossiping can save on transmission cost, but it may propagate stale version information, necessitating transmitter intervention. With imperfect communication links, we formulate an infinite horizon average cost Markov Decision Process (MDP) in order to jointly optimize receivers' VAoIs, sampling and transmission costs in a time slotted fully connected (FC) gossip network. We first solve the MDP using Relative Value Iteration (RVI) algorithm. Then, we prove that the sampling decision has a \emph{threshold structure} in the transmitter VAoI. Further, under reliable gossiping links, we show that transmission and idling has an \emph{VAoI-difference threshold structure with minimum VAoI activation}. Numerical results obtained using RVI show that the optimal policy consistently outperforms natural baselines. These results characterize how source sampling, direct transmission, and receiver gossip
should be jointly coordinated for cost-aware semantic freshness in gossip
networks.
\end{abstract}


\section{Introduction}
\IEEEPARstart{I}{nformation} freshness is central to robust and dependable
performance in status-update systems such as sensor networks
\cite{ZhengTWC2024}, connected and autonomous vehicles (CAVs)
\cite{ChenOJCOMS2025}, and Internet-of-Things (IoT) networks
\cite{KahramanJIOT2024}, among others. Timely information updates are crucial
for control, estimation, and decision making performance
\cite{SoleymaniCUP2023}. Traditionally, information freshness is measured using
Age of Information (AoI) \cite{AoI}, which has become the most widely used
metric for tracking timeliness in such systems \cite{AoISurvey}. However, AoI
measures freshness through elapsed time after update generation, and therefore
may penalize staleness even when the source content has not changed. This
limitation is increasingly important as networked applications move toward
goal-oriented and semantic communication paradigms \cite{SemCom}, with 6G
systems expected to connect massive numbers of intelligent devices. Semantic
communication calls for freshness metrics that reflect meaningful source
changes, motivating a shift from purely time driven freshness measures
\cite{UysalSemantic2022}.

Semantic communication requires freshness metrics that capture meaningful
content changes, since elapsed time metrics such as AoI are often insufficient
from a semantic perspective. Version Age of Information (VAoI), introduced in
\cite{AoG}, provides such a semantic freshness measure by tracking the number
of meaningful source versions missed by a receiver. Equivalently, VAoI
quantifies freshness through the receiver's version lag relative to the source
\cite{SalimnejadTCOMM2025}. VAoI generalizes the Binary Freshness Metric (BFM)
\cite{BFM} and has been widely used in the freshness analysis of gossip
networks.
\IEEEpubidadjcol

Networked communication and control systems, and status-update systems in
general, often contain local receiver-to-receiver communication links. These
links create \emph{gossiping} opportunities, allowing receivers to exchange
information without requiring every update to pass through a central
transmitter. Gossiping reduces dependence on direct transmissions from the
transmitter because receivers can spread fresh versions among themselves.
However, gossiping can only circulate information already available at the
receivers. Thus, when the receiver network becomes stale, transmitter
intervention is necessary. The transmitter can directly inject information into
the receiver network by paying a transmission cost. If the transmitter itself is
stale, it can sample the source to refresh its own version, but this incurs a
sampling cost. When both the transmitter and the receiver network are relatively
fresh, the transmitter may instead idle and allow receiver gossip to reduce
local VAoI disparities without incurring a direct transmission cost.

In this work, we study a VAoI-based fully connected (FC) gossip network in
which the transmitter observes source version changes but must decide when
acquiring the current source version is worth the sampling cost. The transmitter
can directly transmit its available version into the receiver network by paying a
transmission cost, or it can idle and allow receivers to propagate information
among themselves. This creates a coupled control problem in which sampling,
direct transmission, and idling cannot be optimized independently. We formulate
this problem as an infinite horizon average-cost Markov Decision Process (MDP)
\cite{ZakeriTCOMM2025}. We prove that the optimal sampling decision has a
\emph{threshold structure} in the transmitter VAoI. We characterize the structure of direct transmission and idling under
symmetric links, and prove a \emph{VAoI-difference threshold structure with
minimum-VAoI activation} policy. Numerical results validate the structural findings and compare the optimal
policy against several baselines. The results show that jointly optimizing
sampling, direct transmission, and idling provides significant gains over
policies that use these mechanisms in a fixed or decoupled manner.
\subsection{Related Works}

Before the introduction of VAoI in gossip networks \cite{AoG}, AoI had already been considered in that area \cite{AgeOfGossip2013}. Subsequent works have shown that the topology of how nodes are connected in the network plays a central role in the freshness of any gossip network. Hierarchical organization, such as clustering of nodes, also impacts freshness, as shown in \cite{VAoIClustered}, where the grouping of receivers and the hierarchical structure affect the freshness scaling behavior. Further, rate-changing gossip in clustered networks is shown to have an effect on freshness as shown in \cite{RCGossip}. Network topology and other topological effects on freshness have been investigated in \cite{AgeGrid,RandomBipartiteGossip2024,DynamicGossip2025}.The results indicate that network structure and connectivity among the nodes impact the dissemination of fresh information, and can be considered key aspects of freshness.

Recent studies have considered other viewpoints of gossiping. Coded update dissemination is analyzed in \cite{CodedGossip2025}, where memory and memoryless coding schemes were shown to influence the age of updates. Push-pull gossip protocols are investigated in \cite{PushPullGossip2025}, providing an additional alternative mechanism through which nodes can exchange information. Mobility of nodes has been studied in \cite{SrivastavaMobility,VAoIContact}, where node mobility and contact opportunities affect the flow of fresh updates. A survey of closely related works can be found in \cite{AoGSurvey}.

These works offer insight into information freshness in gossip networks. They primarily analyze systems with specific topologies, gossiping dynamics, and processes. They do not directly address how to optimally control freshness in the network. Policy optimization for freshness in gossip networks has only recently begun to receive attention. In \cite{MitraAllerton2022}, age-aware multiple access is studied, showing that update scheduling can be designed with freshness considerations. Opportunistic gossiping in dense networks is investigated in \cite{MitraINFOCOM2023}, where the update process is adapted to exploit favorable dissemination opportunities. A learning based approach for improving fairness in sparse gossip networks is proposed in \cite{MitraICMLCN2024}, highlighting the role of adaptive policies. More closely related to our setting, \cite{VAoIPolicy} formulates a VAoI optimization problem in a \emph{unidirectional ring network} with an energy-harvesting sensor and an aggregator, and shows that the optimal update policy has a threshold structure. Our prior works \cite{TransmitIdleICC2026, TransmitIdleITW2026} study versions of the problem for two gossiping receivers. These works show that freshness in gossip networks can be improved not only by changing topology or gossip rates, but also by formulating and optimizing a control policy.

Outside the context of gossip networks, the structure of optimal policies for status updating has been well studied. Prior works have characterized policy structures in routing, energy harvesting, wireless broadcasting, and federated learning \cite{UysalMobiHoc2025,ZakeriTWC2024,SoleymaniTIT2024,AbdElmagidJoint2020,MaatoukAoII2022,ArafaICC2018,ArafaTIT2020,ArafaTWC2019,YangICASSP2020,ArafaTGCN2022,ArafaTCOMM2021,HsuTMC2020,ZhouTCOMM2019}. These results show that optimal freshness control often has exploitable structure. Such structural characterizations provide analytical insight into why certain update actions are optimal in different regions of the state space. Additionally, structural results help reduce the effective complexity of policy search by replacing the full search space with a smaller family of structured policies.

Beyond the analytical simplification that structural analysis can offer, these properties are useful for learning-based control. Reinforcement learning methods for large MDPs can suffer from high sample complexity. Structure-aware learning methods address this issue by embedding structure into the learning architecture and policy class \cite{NeurIPS2022,BuraTON2026,RoyTGCN2025,ThresholdNW1,ThresholdNW2,StructuredRL,JitaniTCCN2022}, thereby ensuring faster and more accurate learning. These works motivate the study of structural optimality as a practical tool for scalable policy computation, which can speed up learning and make online and real-time implementation more practical.

\subsection{Contributions}

Gossiping can significantly improve freshness by allowing receivers to exchange
locally available information. However, gossiping alone cannot create new source
versions. It can only redistribute versions that are already present in the
receiver network. Thus, when the receiver network becomes stale, transmitter
intervention is necessary. The transmitter must decide whether to sample the
source (with a sampling cost), transmit its available version to a receiver (with a transmission cost), or idle and allow receiver
gossip to reduce local VAoI disparities. This creates a fundamental control
tradeoff between semantic freshness and operation costs. In this paper, we study this tradeoff under the Version Age of Information
(VAoI) metric. Unlike models in which the transmitter always has the freshest
source version, we explicitly account for transmitter staleness and costly
source sampling. The resulting problem asks:
\begin{quote}
\emph{When to sample the source, whom to serve directly, and when to idle so
that gossip can improve freshness in a cost-effective manner?}
\end{quote}
The main contributions are summarized as follows.

\begin{itemize}
    \item We formulate a VAoI-based freshness control problem for a general
    $n$-receiver fully connected (FC) gossip network with costly source
    sampling and lossy direct transmission.

    \item We cast the problem as an infinite-horizon average-cost Markov
    Decision Process (MDP), where the objective is to minimize the long-term receiver VAoI together with sampling and transmission costs. We
    compute the optimal policy using Relative Value Iteration (RVI).

    \item We prove that the optimal sampling decision has a threshold structure
    in the transmitter VAoI. 

    \item We characterize the structure of the transmission and idling actions under
    symmetric links. We prove that an optimal direct transmission targets a
    receiver with the largest VAoI, and we establish a VAoI-difference threshold
    structure for the transmission versus idling decision.

    \item We numerically evaluate the optimal policy against natural baseline
    policies to
    illustrate the threshold structures and show the benefit of jointly
    optimizing sampling, direct transmission, and idling.
\end{itemize}
\section{System Model}

\begin{figure}[t]
    \centering
    \includegraphics[width=0.75\linewidth]{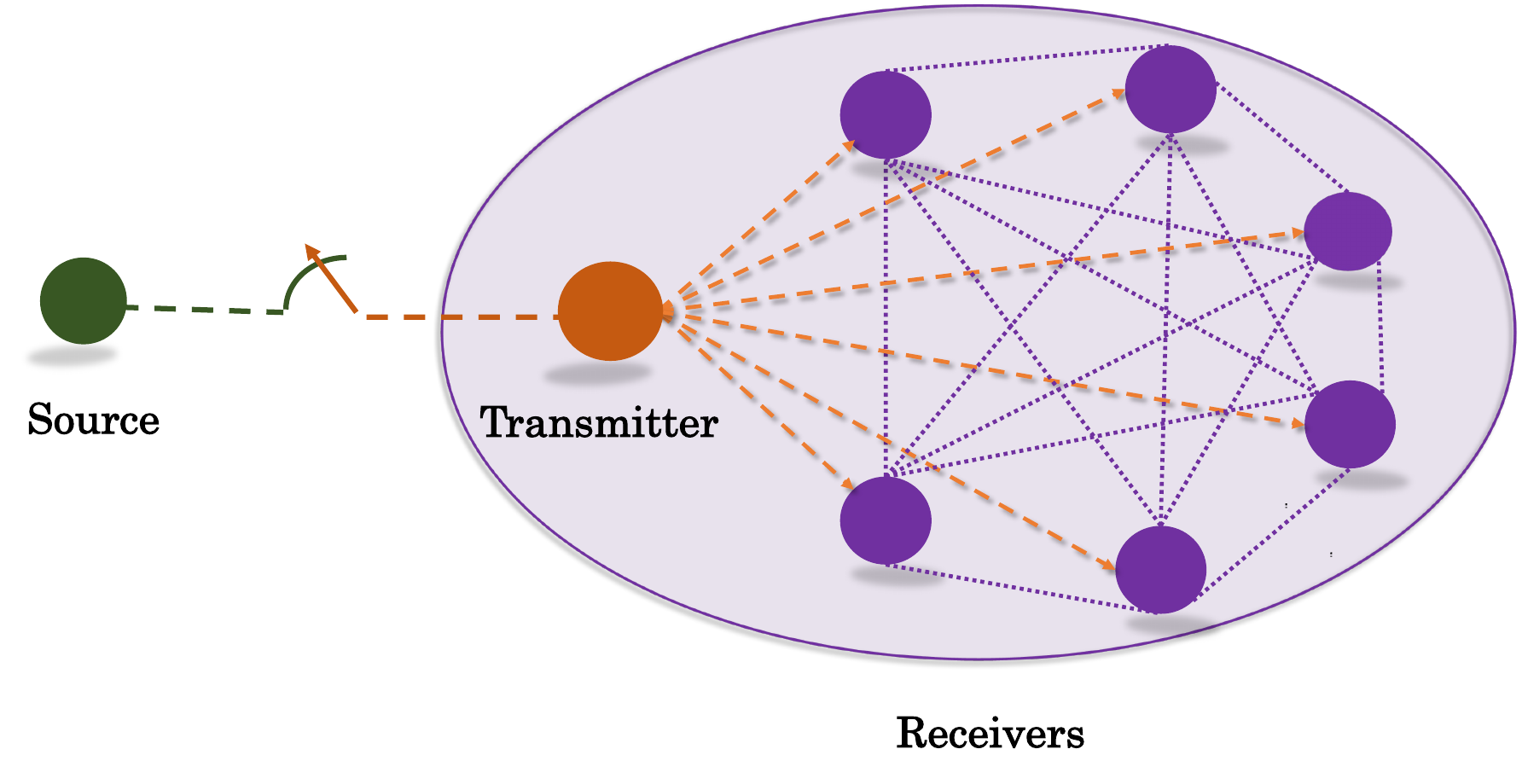}
    \caption{System model showing a source, transmitter and a fully connected (FC) gossip network with $n=6$.}
    \label{fig:system_model}
\end{figure}

We consider a time-slotted system indexed by $t\in\{0,1,2,\ldots\}$, consisting of $n$ fully connected (FC) gossiping receivers, denoted by RX$_1,\ldots,$ RX$_n$. The source evolves over time and generates new information. A system with $n=6$ is shown in Fig. \ref{fig:system_model}. Let $X(t)$ denote the source information at slot $t$. The source content, which we denote as the \emph{version} of the information, remains unchanged from one slot to the next with probability $1-\alpha$ and changes with probability $\alpha$, independently across slots. Let $\delta\in\{0,1\}$ denote the source version increment in a slot, where
\begin{align}
\mathbb P(\delta=1)=\alpha,
\qquad
\mathbb P(\delta=0)=1-\alpha .
\end{align}
Here, $\delta=1$ corresponds to the generation of a new source version, while $\delta=0$ corresponds to no source version generation at the beginning of a slot. The transmitter is aware of source version changes, and can therefore track the version ages of all nodes \cite{VAoIPolicy}. However, observing that the source version has changed, does not give the transmitter the current source content. To obtain the current information version, the transmitter has to sample and pay a sampling cost. The transmitter can transmit its available version to one receiver, or remain idle and allow receivers to exchange their available versions through gossiping.

We assume that receivers in the FC gossip network send ACK/NACK feedback after each direct transmission attempt from the transmitter or any successful gossiping between them. If a receiver obtains multiple successful gossip updates in the same slot, it keeps the freshest received version, i.e., the one with the minimum VAoI, and gives feedback using ACK to the transmitter about its change in VAoI. Therefore, the transmitter can track the receiver VAoIs, and has perfect knowledge of the system state, which we describe in the following section.

\subsection{System State}

Let $N_s(t)$ denote the source version at the beginning of slot $t$, let $N_{tx}(t)$ denote the source version at the transmitter, and let $N_i(t)$ denote the source version at receiver $i$. The version age of $x\in\{tx,1,\ldots,n\}$ is defined as
\begin{align}
v_x(t)\triangleq N_s(t)-N_x(t).
\end{align}
Accordingly, $v_x(t)=0$ means that node $x$ has the current source version, while $v_x(t)>0$ means that node $x$ is behind the current source information version by $v_x(t)$ versions.
The system state is the VAoI of the transmitter and the $n$ receiver nodes:
\begin{align}
s(t)=\big(v_{tx}(t),v_1(t),\ldots,v_n(t)\big).
\end{align}

Since every receiver information version comes from the transmitter or from another receiver that received information from the transmitter, the transmitter cannot be ``older'' than any receiver. Hence, the state space satisfies $ 0\le v_{tx}(t)\le v_i(t),~\forall i \in \{1,\ldots,n\}$. For numerical computation, we use a finite truncation parameter $V_{\max}$ and define the state space as:

\begin{align}
\mathcal S
=
\{ (v_{tx},v_1,\ldots,v_n):
0\le v_{tx}\le v_i&\le V_{\max},\nonumber\\ 
& i=1,\ldots,n\}.
\end{align}

At the beginning of each slot, when the system is in state $s(t)=(v_{tx},v_1,\ldots,v_n)$, the transmitter chooses an action based on the state. During the slot, the
source may generate a new information version. 
After the action is selected, the source evolution is realized, producing the
intermediate VAoIs:
\begin{align}
    \tilde v_{tx}
=
(v_{tx}+\delta)\wedge V_{\max},
\label{eq:intermediate_vaoi1}
\end{align}
\begin{align}
\tilde v_i
=
(v_i+\delta)\wedge V_{\max},
\qquad i=1,\ldots,n,
\label{eq:intermediate_vaoi2}
\end{align}
where $a\wedge b=\min\{a,b\}$. The system state after the source evolution becomes
\begin{align}
    \tilde s(t)
=
\big(\tilde v_{tx},\tilde v_1,\ldots,\tilde v_n\big).
\end{align}
The chosen action is applied to this intermediate state. Each
time slot $t$ proceeds as follows. First, the transmitter observes the current
state $s(t)$ and selects an action $u(t)\in\mathcal A$. Next, the source
evolution $\delta(t)\in\{0,1\}$ is realized, producing the intermediate VAoIs
$(\tilde v_{tx},\tilde v_1,\ldots,\tilde v_n)$. The chosen action $u(t)$ is
then applied to this intermediate state. Finally, communication links' reliability outcomes determine the next state $s(t+1)$ and
ACK/NACK feedbacks are sent.

\subsection{Actions and Transitions}
The transmitter chooses one action from the action set at the beginning of each time slot: 
\begin{align}
\mathcal A=\{0,1,\ldots,n,n+1\}.
\end{align}
The actions are defined as follows:
\begin{align}
u(t)=
\begin{cases}
0, & \text{idle},\\[1mm]
i, & \text{transmit to receiver } i,\quad i\in\{1,\ldots,n\},\\[1mm]
n+1, & \text{sample source}.
\end{cases}
\end{align}

If the transmitter chooses action $u(t)=i$, where $i\in\{1,\ldots,n\}$, then it attempts to send its currently available version to receiver $i$. The direct link to receiver $i$ succeeds with probability $p_i$ and fails with probability $1-p_i$. If the transmission succeeds, receiver $i$ adopts the transmitter's version. Therefore, its VAoI after the current slot becomes $v_i(t+1)=\tilde v_{tx}$.
All other receivers remain at their intermediate VAoIs. Thus, under action $u(t)=i$,
\begin{align}
s(t+1)=
\begin{cases}
(\tilde v_{tx},\tilde v_1,\ldots,\tilde v_{i-1},
\tilde v_{tx},\tilde v_{i+1},\ldots,\tilde v_n),
\quad \text{w.p. } p_i,\\[1mm]
(\tilde v_{tx},\tilde v_1,\ldots,\tilde v_n),
\quad \text{w.p. } 1-p_i .
\end{cases}
\end{align}
A successful transmission to receiver $i$ resets it to the transmitter VAoI $\tilde v_{tx}$. Receiver $i$ becomes fully fresh only when the transmitter itself is fresh, i.e., when $\tilde v_{tx}=0$. We assume that direct transmission and RX gossip share the same communication resources. Hence, gossip occurs only in slots without a direct transmission. Only when the transmitter idles or samples, the receivers can gossip. When this is the case, in those time slots, each receiver can transmit to one other receiver. However, a receiver may receive gossip from multiple other receivers in the same slot. Let
$r_j(t)\in\{1,\ldots,n\}\setminus\{j\}$
denote the receiver selected by receiver $j$ as its gossip destination. Each receiver selects its target uniformly at random:
\begin{align}
\mathbb P(r_j(t)=i)
=
\frac{1}{n-1},
\qquad i\neq j.
\end{align}
The gossip attempt from $j$ to $i$ succeeds with probability $p_{v,ji}$, where
$0<p_{v,ji}<1$, $\forall$ $j$. Let $G_{ji}(t)$ denote the event that receiver $j$ successfully gossips its version to receiver $i$ in slot $t$. For each receiver $i$, the set of successful incoming gossip senders is
\begin{align}
    \mathcal G_i(t)
=
\{j\neq i:
r_j(t)=i
\text{ and }G_{ji}(t)\text{ occurs}\}.
\end{align}
Receiver $i$ keeps the freshest version among all successfully received gossip versions and its own information version. Since smaller VAoI means fresher information, the receiver update rule is:
\begin{align}
    v_i(t+1)
=
\min
\left\{
\tilde v_i,
\min_{j\in \mathcal G_i(t)} \tilde v_j
\right\},
\qquad i=1,\ldots,n,
\label{eq:gossip_individual}
\end{align}
Therefore, if receiver $i$ receives no successful gossip packet, then $v_i(t+1)=\tilde v_i$. 

Under the idle action $u(t)=0$, the transmitter does not sample, and hence $v_{tx}(t+1)=\tilde v_{tx}$.
The receiver VAoIs change based on gossiping. Therefore,using \eqref{eq:gossip_individual}, the state becomes:
\begin{align}
    s(t+1)
=
\Big(
\tilde v_{tx},
\min\{\tilde v_1,\min_{j\in\mathcal G_1(t)}&\tilde v_j\},
\ldots,\nonumber\\
&\min\{\tilde v_n,\min_{j\in\mathcal G_n(t)}\tilde v_j\}
\Big).
\end{align}

If the transmitter chooses the sampling action $u(t)=n+1$, the transmitter VAoI resets to zero,
$v_{tx}(t+1)=0$. As stated previously, during the same slot, receivers are allowed to gossip. Thus, under sampling using \eqref{eq:gossip_individual}, the next state is
\begin{align}
s(t+1)
=
\Big(
0,
\min\{\tilde v_1,\min_{j\in\mathcal G_1(t)}&\tilde v_j\},
\ldots,\nonumber\\
&\min\{\tilde v_n,\min_{j\in\mathcal G_n(t)}\tilde v_j\}
\Big).
\end{align}

Receivers benefit from sampling in later time slots, after the transmitter subsequently transmits its newly sampled version, or through gossiping after that information version enters the network.

\subsection{Costs}
The costs partly depend on the transmitter action. Let $C_s \ge 0 $ denote the sampling cost and $C_{\mathrm{tx}} \ge 0$ denote the cost of transmission. The action-dependent part of the cost is
\begin{align}
C(u)=
\begin{cases}
0, & u=0,\\
C_{\mathrm{tx}}, & u\in\{1,\ldots,n\},\\
C_s, & u=n+1.
\end{cases}
\end{align}

There is no cost incurred when idling ($u=0$). The transmitter uses the channel statistics knowledge (i.e., $p_i$'s, and $p_{ij}$'s) and defines the system cost at a given time slot as the expected sum of all RXs' VAoIs at the next time slot plus the action dependent cost (if any).
Therefore, for state $s$ and action $u$, the immediate cost is:
\begin{align}
c(s,u)
=
\mathbb E
\left[
\sum_{i=1}^n v_i(t+1)
\,\middle|\,
s(t)=s,\ u(t)=u
\right]
+
C(u).
\label{eq:cost}
\end{align}

\subsection{Problem Formulation}

We now formulate the infinite-horizon average cost MDP. Our objective is to find a stationary policy $\pi:\mathcal S\to\mathcal A$ that maps each state to an action and optimizes the long term average system cost. For a given policy $\pi$, the long-term average cost starting from $s(0)=s$ is

\begin{align}
\rho^\pi(s)
=
\limsup_{T\to\infty}
\frac{1}{T}
\mathbb E^\pi
\left[
\sum_{t=0}^{T-1}
c(s(t),\pi(s(t)))
\,\middle|\,
s(0)=s
\right].
\end{align}
The objective is to find a stationary policy minimizing this cost, i.e., to evaluate:
\begin{align}
    \pi^\star\in\arg\min_\pi \rho^\pi(s).
\end{align}

In subsequent sections, we show that the feasible state space is closed and
that, under $0<\alpha<1$, $0<p_i<1$, $\forall i$, and $0<p_{v,ji}<1$,
$\forall i$ and $\forall j\neq i$, the MDP is communicating. Therefore, by
standard finite-state average-cost MDP results, the optimal average cost is
independent of the initial feasible state and the average-cost optimality
equation admits a solution \cite{Bertsekas}. Let $\rho^\star$ denote the optimal average cost
and let $U(s)$ denote the relative value function. The average-cost Bellman
optimality equation is
\begin{align}
\rho^\star+U(s)
=
\min_{u\in\mathcal A}
\Bigg\{
c(s,u)
+
\mathbb E
\left[
U(s')
\,\middle|\,
s,u
\right]
\Bigg\},
\quad s\in\mathcal S,
\end{align}
where $s'$ denotes the next state induced by the source evolution, transmission
outcome, and gossiping outcome under the state-action pair $(s,u)$. We define the action value as
\begin{align}
Q(s,u)
\triangleq
c(s,u)
+
\mathbb E
\left[
U(s')
\,\middle|\,
s,u
\right].
\label{eq:qval}
\end{align}
Thus, an optimal stationary deterministic policy is obtained by
\begin{align}
\pi^\star(s)\in\arg\min_{u\in\mathcal A}Q(s,u).
\end{align}

We compute the optimal policy numerically using relative value iteration (RVI)\cite{Puterman}, according to the steps detailed in Algorithm \ref{alg:rvi}
and analyze its structural properties in the following sections.

\section{Properties of the Markov Decision Process}
We first establish basic properties of the MDP for completeness. The first property shows that the state space is closed under all transition outcomes. The second property shows that the MDP is communicating. For two states $s=(v_{tx},v_1,\ldots,v_n)$ and
$s'=(v'_{tx},v'_1,\ldots,v'_n)$, we write $s\preceq s'$ if
\begin{align}
    v_{tx}\le v'_{tx},
\qquad
v_i\le v'_i,\quad i=1,\ldots,n.
\label{eq:sles'}
\end{align}

\begin{proposition}[Closure]
\label{prop:feasibility}
If $s(t)\in\mathcal S$, then $s(t+1)\in\mathcal S$, $\forall u\in\mathcal A$ and for all transition outcomes. Hence, the state space $\mathcal S$ is closed.
\end{proposition}

\begin{proof}
See Appendix~\ref{app:feasibility}.
\end{proof}

\begin{proposition}[Communication]
\label{prop:weak_comm}
Assume $0<\alpha<1$, $0<p_i<1$, $\forall i\in\{1,\ldots,n\}$, and
$0<p_{v,ji}<1$, $\forall j\neq i$. Then the MDP on $\mathcal S$ is communicating.
\end{proposition}

\begin{proof}
See Appendix~\ref{app:weak_comm}.
\end{proof}

We next establish the properties of the relative value function that are used to prove the structure.

\begin{algorithm}[t]
\caption{Relative Value Iteration}
\label{alg:rvi}
\footnotesize
\begin{algorithmic}[1]
\STATE \textbf{Initialize:} $s_{\rm ref}$, $\epsilon$, $k=0$, and
$U^{(0)}(s)=0$ for all $s\in\mathcal S$.
\REPEAT
    \FOR{$s\in\mathcal S$}
        \STATE
        $
        \bar U^{(k+1)}(s)
        =
        \min\limits_{u\in\mathcal A}
        \left[
        c(s,u)+
        \sum\limits_{s'\in\mathcal S}
        P(s'|s,u)U^{(k)}(s')
        \right].
        $
        \STATE
        $
        U^{(k+1)}(s)=
        \bar U^{(k+1)}(s)-\bar U^{(k+1)}(s_{\rm ref}).
        $
    \ENDFOR
    \STATE $k\leftarrow k+1$.
\UNTIL{$\max\limits_{s\in\mathcal S}
\left|U^{(k)}(s)-U^{(k-1)}(s)\right|<\epsilon$}
\STATE Set $\rho^{(k)}=\bar U^{(k)}(s_{\rm ref})$.
\STATE Generate the deterministic policy
\STATE
$
\pi^\star(s)
\in
\arg\min\limits_{u\in\mathcal A}
\left[
c(s,u)+
\sum\limits_{s'\in\mathcal S}
P(s'|s,u)U^{(k)}(s')
\right],
\quad \forall s\in\mathcal S.
$
\STATE \textbf{Output:} $\pi^\star$, $U^{(k)}$, and $\rho^{(k)}$.
\end{algorithmic}
\end{algorithm}

\section{Monotonicity of the Relative Value Function}

We establish monotonicity property of the relative value function. For a fixed action \(u\), source increment \(\delta\), and transition outcome
\(\omega\), let
\begin{align}
F_{u,\omega,\delta}(s),
\qquad
F_{u,\omega,\delta}(s')
\end{align}
denote the next states generated from $s$ and $s'$ under the
same realization where $s \preceq s'$ according to \eqref{eq:sles'}.

\begin{lemma}[Order preservation]
\label{lem:order_preservation}
For every fixed $u\in\mathcal A$, $\delta\in\{0,1\}$, and transition outcome
$\omega$, the next-state map is order-preserving, which means:
\begin{align}
s\preceq s'
\quad\Longrightarrow\quad
F_{u,\omega,\delta}(s)
\preceq
F_{u,\omega,\delta}(s').
\end{align}
\end{lemma}

\begin{proof}
See Appendix~\ref{app:order_preservation}.
\end{proof}

\begin{remark}
The mapping $F_{u,\omega,\delta}(\cdot)$ denotes the next state 
under a fixed action $u$, fixed source evolution $\delta$, and fixed realization of all link outcomes and gossip target selections, denoted by
$\omega$. 
\end{remark}

\begin{lemma}[Cost monotonicity]
\label{lem:cost_monotonicity}
If \(s\preceq s'\), then for every action \(u\in\mathcal A\),
$$
c(s,u)\le c(s',u).
$$
\end{lemma}

\begin{proof}
See Appendix~\ref{app:cost_monotonicity}.
\end{proof}

\begin{proposition}[Relative Value Function Monotonicity]
\label{prop:rvf_monotonicity}
The relative value function $U(s)$ is nondecreasing with respect to the state:
\begin{align}
s\preceq s'
\quad\Longrightarrow\quad
U(s)\le U(s').
\end{align}
\end{proposition}

\begin{proof}
We prove the result using inductive hypothesis. Let $U^{(k)}$ denote the
$k$-th RVI iterate and initialize $U^{(0)}(s)=0$, $\forall s\in\mathcal S$.
Thus, $U^{(0)}$ is nondecreasing. Assume $U^{(k)}$ is nondecreasing. The RVI update is
\begin{align}
U^{(k+1)}(s)
=
T U^{(k)}(s)
-
T U^{(k)}(s_{\mathrm{ref}}),
\label{eq:rviupdate}
\end{align}
where $s_{\mathrm{ref}}$ is fixed and
\begin{align}
T U^{(k)}(s)
=
\min_{u\in\mathcal A}
\left\{
c(s,u)
+
\mathbb E
\left[
U^{(k)}(S)
\,\middle|\,
s,u
\right]
\right\}.
\label{eq:TUdef}
\end{align}
The term $T U^{(k)}(s_{\mathrm{ref}})$ is independent of $s$, so
it does not affect monotonicity.

Now let $s\preceq s'$ and fix an action $u\in\mathcal A$. For a fixed source
increment $\delta$ and transition outcome $\omega$, define
$S=F_{u,\omega,\delta}(s)$ and $S'=F_{u,\omega,\delta}(s')$. By
Lemma~\ref{lem:order_preservation},
\begin{align}
S\preceq S'.
\end{align}
Since $U^{(k)}$ is nondecreasing, we have
\begin{align}
U^{(k)}(S)\le U^{(k)}(S').
\end{align}
This holds for every realization of $\delta$ and $\omega$. Therefore, taking
expectation over all transition outcomes gives
\begin{align}
\mathbb E
\left[
U^{(k)}(S)
\,\middle|\,
s,u
\right]
\le
\mathbb E
\left[
U^{(k)}(S')
\,\middle|\,
s',u
\right].
\end{align}
By Lemma~\ref{lem:cost_monotonicity}, $c(s,u)\le c(s',u)$. Combining this with
the previous inequality gives
\begin{align}
&c(s,u)
+
\mathbb E
\left[
U^{(k)}(S)
\,\middle|\,
s,u
\right] \nonumber\\
&\qquad \le
c(s',u)
+
\mathbb E
\left[
U^{(k)}(S')
\,\middle|\,
s',u
\right].
\end{align}
Since this inequality holds for every $u\in\mathcal A$, taking the minimum over
actions gives using \eqref{eq:TUdef},
\begin{align}
T U^{(k)}(s)
\le
T U^{(k)}(s').
\end{align}
Therefore, by \eqref{eq:rviupdate},
\begin{align}
U^{(k+1)}(s)
&=
T U^{(k)}(s)
-
T U^{(k)}(s_{\mathrm{ref}}) \nonumber\\
&\le
T U^{(k)}(s')
-
T U^{(k)}(s_{\mathrm{ref}}) \nonumber\\
&=
U^{(k+1)}(s').
\end{align}
Thus, $U^{(k+1)}$ is nondecreasing. By induction, every RVI iterate is
nondecreasing. Letting $k\to\infty$, the normalized RVI iterates converge to
the relative value function $U$. Hence,
\begin{align}
U(s)
=
\lim_{k\to\infty}U^{(k)}(s)
\le
\lim_{k\to\infty}U^{(k)}(s')
=
U(s').
\end{align}
Therefore, $s\preceq s'$ implies $U(s)\le U(s')$.
\end{proof}

\section{Sampling Threshold of the Optimal Policy}

We first establish that the optimal sampling decision has a threshold structure
in the transmitter VAoI $v_{tx}$. For fixed receiver VAoIs, sampling becomes
more attractive as the transmitter becomes more stale. In particular, once
sampling is optimal at some $v_{tx}$, it remains optimal for all larger
feasible transmitter VAoIs.

\begin{theorem}[Sampling Threshold]
\label{thm:sampling_threshold}
For fixed receiver VAoIs, the optimal sampling decision is threshold-based in
the transmitter VAoI $v_{tx}$. Equivalently, if sampling is optimal at some
feasible $v_{tx}$, then it is also optimal for every larger feasible transmitter
VAoI with the same receiver VAoIs.
\end{theorem}

\begin{proof}
Fix receiver VAoIs $(v_1,\ldots,v_n)$ and consider two feasible states
$s=(v_{tx},v_1,\ldots,v_n)$ and
$s'=(v'_{tx},v_1,\ldots,v_n)$ with $v_{tx}\le v'_{tx}$, which gives $s\preceq s'$.

First, we consider the sampling action $u=n+1$. Under sampling, the transmitter
VAoI is reset to zero ($v_{tx} = 0$ and $v_{tx'} = 0 $). Since the receiver VAoIs are identical in $s$ and $s'$ ($v_i = v_i'$ for all $i \in \{1,2,\ldots ,n\}$),
the same source evolution and gossip outcome produce identical next states.
Thus,
\begin{align}
F_{n+1,\omega,\delta}(s)
=
F_{n+1,\omega,\delta}(s'),
\qquad \forall \delta,\omega .
\end{align}
Hence,
\begin{align}
U\!\left(F_{n+1,\omega,\delta}(s)\right)
=
U\!\left(F_{n+1,\omega,\delta}(s')\right),
\qquad \forall \delta,\omega .
\end{align}
Taking expectation over all transition outcomes gives
\begin{align}
\mathbb E
\left[
U\!\left(F_{n+1,\omega,\delta}(s)\right)
\,\middle|\,
s,n+1
\right]
=
\mathbb E
\left[
U\!\left(F_{n+1,\omega,\delta}(s')\right)
\,\middle|\,
s',n+1
\right].
\end{align}
Moreover, the immediate costs are identical because the expected next receiver
VAoI sum is the same in both states and the sampling cost is $C_s$ in both
cases. Hence by \eqref{eq:cost} ,
\begin{align}
c(s,n+1)=c(s',n+1).
\end{align}
Using the definition of $Q(s,u)$ from \eqref{eq:qval}, we obtain
\begin{align}
Q(s,n+1)=Q(s',n+1).
\label{eq:samp}
\end{align}

Now consider any non-sampling action $u\in\{0,1,\ldots,n\}$. By
Lemma~\ref{lem:cost_monotonicity}, $c(s,u)\le c(s',u)$. Also, by
Lemma~\ref{lem:order_preservation} and
Proposition~\ref{prop:rvf_monotonicity},
\begin{align}
U\!\left(F_{u,\omega,\delta}(s)\right)
\le
U\!\left(F_{u,\omega,\delta}(s')\right),
\qquad \forall \delta,\omega .
\end{align}
Taking expectation over all transition outcomes gives
\begin{align}
\mathbb E
\left[
U\!\left(F_{u,\omega,\delta}(s)\right)
\,\middle|\,
s,u
\right]
\le
\mathbb E
\left[
U\!\left(F_{u,\omega,\delta}(s')\right)
\,\middle|\,
s',u
\right].
\end{align}
Combining this with $c(s,u)\le c(s',u)$ and using \eqref{eq:qval} gives:
\begin{align}
Q(s,u)\le Q(s',u),
\qquad u\in\{0,1,\ldots,n\}.
\label{eq:non-samp}
\end{align}
Hence,
\begin{align}
\min_{u\in\{0,1,\ldots,n\}}Q(s,u)
\le
\min_{u\in\{0,1,\ldots,n\}}Q(s',u).
\end{align}

For fixed receiver VAoIs, we define the sampling gap as a function of the
transmitter VAoI, $v_{tx}$:
\begin{align}
\begin{aligned}
\Delta(v_{tx})
&\triangleq
Q\big((v_{tx},v_1,\ldots,v_n),n+1\big) \\
&\quad -
\min_{u\in\{0,1,\ldots,n\}}
Q\big((v_{tx},v_1,\ldots,v_n),u\big).
\end{aligned}
\end{align}
By the preceding arguments in \eqref{eq:samp}, the first term in $\Delta(v_{tx})$ is independent
of $v_{tx}$, while by \eqref{eq:non-samp} the second term is nondecreasing in $v_{tx}$. Therefore,
$\Delta(v_{tx})$ is nonincreasing in $v_{tx}$.

If sampling is optimal at $s$, then $\Delta(v_{tx})\le0$. Since
$v'_{tx}\ge v_{tx}$ and $\Delta(\cdot)$ is nonincreasing in $v_{tx}$,
\begin{align}
\Delta(v'_{tx})
\le
\Delta(v_{tx})
\le 0.
\end{align}
Thus,
\begin{align}
Q(s',n+1)
\le
\min_{u\in\{0,1,\ldots,n\}}Q(s',u),
\end{align}
so sampling is optimal at $s'$. This proves the threshold property.
\end{proof}

\begin{remark}
Theorem~\ref{thm:sampling_threshold} shows that, for every fixed receiver-VAoI
vector $(v_1,\ldots,v_n)$, there exists a threshold
$\tau(v_1,\ldots,v_n)$ such that sampling is optimal for all feasible
$v_{tx}\ge \tau(v_1,\ldots,v_n)$.
\end{remark}

In states below the sampling threshold, the relevant decision is between direct
transmission and idling. We study these decisions next.
\section{Optimal Policy for Transmission and Idling}

We now turn to the structure of the actions in the states below the sampling threshold. Once sampling is not
selected, the transmitter has to decide whether to idle and allow gossip, or transmit its available versions to receivers. To obtain clean
structural comparisons, we focus on the
symmetric setting, where all direct transmission links have same success probability and all gossip links have the same success probability. With these assumptions, for the rest of the paper, we assume
$
p_i=p, \forall i
$
and
$
p_{v,ji}=p_v, \forall j\neq i.
$

Under symmetric links, receivers are statistically identical apart from their VAoIs. If two states have the same transmitter VAoI $v_{tx}$
and differ only because some receivers have exchanged their VAoI values, then the
two states should have the same relative value function. Mathematically, this exchange is
represented by a permutation of the receiver VAoIs. If
$s=(v_{tx},v_1,\ldots,v_n)$ and $\sigma$ is a permutation of
$\{1,\ldots,n\}$, then the permuted state is defined as
\begin{align}
\sigma(s)
\triangleq
(v_{tx},v_{\sigma^{-1}(1)},\ldots,v_{\sigma^{-1}(n)}).
\end{align}
Thus, in $\sigma(s)$, the receiver in position $i$ has the VAoI that belonged
to receiver $\sigma^{-1}(i)$ in $s$. Equivalently, the VAoI of receiver $j$ in
$s$ is moved to receiver $\sigma(j)$ in $\sigma(s)$. We formalize this
property next.

\begin{figure}[t]
    \centering
    \includegraphics[width=0.75\linewidth]{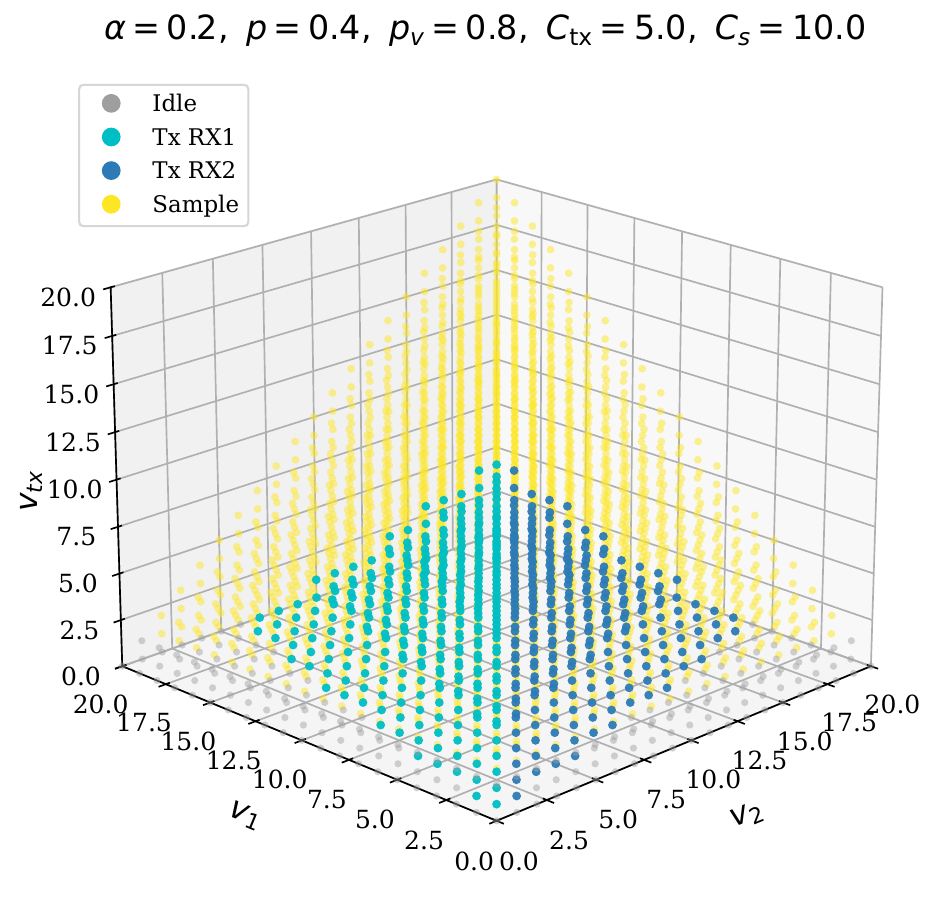}
    \caption{Optimal policy over the three-dimensional VAoI state space
$(v_{tx},v_1,v_2)$ for the two-receiver case $(n=2)$, obtained using RVI.}
    \label{fig:optimal_policy_3d}
\end{figure}

\begin{proposition}[Permutation Invariance]
\label{prop:symmetry}
Under symmetric direct and gossip links, the relative value function is invariant
under any permutation of receiver VAoIs. That is, for any feasible state
$s\in\mathcal S$ and any permutation $\sigma$ of $\{1,\ldots,n\}$,
\begin{align}
U(s)=U(\sigma(s)).
\end{align}
\end{proposition}

\begin{proof}
Let $\sigma$ be any permutation of $\{1,\ldots,n\}$. Since $\sigma$ only
permutes receiver VAoIs, $\sigma(s)\in\mathcal S$ where $s\in\mathcal S$. The corresponding action permutation is $\sigma_{\mathcal A}(0)=0$,
$\sigma_{\mathcal A}(n+1)=n+1$, and
$\sigma_{\mathcal A}(i)=\sigma(i)$ for $i\in\{1,\ldots,n\}$.
Thus, idling and sampling are unchanged, while a direct transmission to RX$_i$
is mapped to a direct transmission to RX$_{\sigma(i)}$.

We prove the result by induction on the RVI iterates. Initialize
$U^{(0)}(s)=0$ for all $s\in\mathcal S$. Then $U^{(0)}$ is permutation
invariant. Assume that $U^{(k)}$ is permutation invariant, i.e.,
\begin{align}
U^{(k)}(s)=U^{(k)}(\sigma(s)),
\qquad \forall s\in\mathcal S .
\end{align}
Similar to \eqref{eq:qval} we have, 
\begin{align}
Q^{(k)}(s,u)
\triangleq
c(s,u)
+
\mathbb E
\left[
U^{(k)}(s')
\,\middle|\,
s,u
\right],
\end{align}
where $s'$ denotes the next state generated from state $s$ under action $u$. Under symmetric direct and gossip links, permuting receiver VAoIs does not
change the immediate cost. Hence, for every $s\in\mathcal S$ and
$u\in\mathcal A$,
\begin{align}
c(s,u)
=
c(\sigma(s),\sigma_{\mathcal A}(u)).
\label{eq:immcost}
\end{align}
Moreover, if $s'$ is the next state generated from $(s,u)$, then
$\sigma(s')$ has the same distribution as the next state generated from
$(\sigma(s),\sigma_{\mathcal A}(u))$. Therefore, using the induction
hypothesis,
\begin{align}
\mathbb E
\left[
U^{(k)}(s')
\,\middle|\,
s,u
\right]
&=
\mathbb E
\left[
U^{(k)}(\sigma(s'))
\,\middle|\,
s,u
\right] \nonumber\\
&=
\mathbb E
\left[
U^{(k)}(s')
\,\middle|\,
\sigma(s),\sigma_{\mathcal A}(u)
\right].
\end{align}
Combining this with \eqref{eq:immcost} gives, 
\begin{align}
Q^{(k)}(s,u)
=
Q^{(k)}(\sigma(s),\sigma_{\mathcal A}(u)).
\end{align}

Since $\sigma_{\mathcal A}$ is a permutation of $\mathcal A$, using the definition from \eqref{eq:TUdef},
\begin{align}
\begin{aligned}
T U^{(k)}(\sigma(s))
&=
\min_{u\in\mathcal A}
Q^{(k)}(\sigma(s),\sigma_{\mathcal A}(u))\\
&=
\min_{u\in\mathcal A}Q^{(k)}(s,u)\\
&=
T U^{(k)}(s).
\end{aligned}
\end{align}

From \eqref{eq:rviupdate}, the RVI update is
\begin{align}
U^{(k+1)}(s)
=
T U^{(k)}(s)
-
T U^{(k)}(s_{\mathrm{ref}}).
\end{align}
Therefore,
\begin{align}
\begin{aligned}
U^{(k+1)}(\sigma(s))
&=
T U^{(k)}(\sigma(s))
-
T U^{(k)}(s_{\mathrm{ref}})\\
&=
T U^{(k)}(s)
-
T U^{(k)}(s_{\mathrm{ref}})\\
&=
U^{(k+1)}(s).
\end{aligned}
\end{align}
Thus, $U^{(k+1)}$ is permutation invariant. By induction, every RVI iterate is
permutation invariant. Taking $k\to\infty$ gives
\begin{align}
U(\sigma(s))=U(s),
\end{align}
for every feasible state $s$ and every receiver permutation $\sigma$.
\end{proof}

We next use this result
together with the monotonicity of \(U\), to compare direct transmissions. It shows that if one receiver has a larger VAoI than others, then transmitting to the larger VAoI is the optimal action.

\begin{proposition}[Serve-Oldest Receiver]
\label{prop:serve_oldest}
For any state $s=(v_{tx},v_1,\ldots,v_n)\in\mathcal S$ and any two receivers
$i,j$ such that $v_i<v_j$, transmitting to RX$_j$ is strictly better than
transmitting to RX$_i$, i.e.,
\begin{align}
Q(s,j)<Q(s,i).
\end{align}
Consequently, among direct transmissions, an optimal direct transmission targets
a receiver in $\arg\max_{\ell\in\{1,\ldots,n\}}v_\ell$.
\end{proposition}

\begin{proof}
Fix a state
\begin{align}
s
=
(v_{tx},v_1,\ldots,v_i,\ldots,v_j,\ldots,v_n),
\end{align}
with receivers $i$ and $j$ satisfying $v_i<v_j$, where WLOG, we assume $i<j$. For a fixed source evolution $\delta$, the intermediate state is
\begin{align}
\tilde s
=
(\tilde v_{tx},\tilde v_1,\ldots,\tilde v_i,\ldots,\tilde v_j,\ldots,\tilde v_n).
\end{align}
Transmissions to RX$_i$ and RX$_j$
succeed with the same probability $p$. If the transmission fails, both actions
lead to the same next state
$s^{\rm fail}(\delta)=\tilde s$. If RX$_i$ is served successfully, the next state is $
s_i^{\rm succ}(\delta) = (\tilde v_{tx},\tilde v_1,\ldots,\tilde v_{tx},\ldots,\tilde v_j,\ldots,\tilde v_n)
$. If RX$_j$ is served successfully, the next state is $s_j^{\rm succ}(\delta)
= (\tilde v_{tx},\tilde v_1,\ldots, \tilde v_i,\ldots,\tilde v_{tx}, \ldots,\tilde v_n).$
Let $\sigma$ be the permutation that swaps receivers $i$ and $j$ and leaves all
other receivers unchanged. Applying $\sigma$ to $s_i^{\rm succ}(\delta)$ gives,
\begin{align}
\sigma(s_i^{\rm succ}(\delta))
=
(\tilde v_{tx},\tilde v_1,\ldots,\tilde v_j,\ldots,\tilde v_{tx},\ldots,\tilde v_n).
\end{align}
Thus, $s_j^{\rm succ}(\delta)$ and
$\sigma(s_i^{\rm succ}(\delta))$ are identical in every coordinate except
RX$_i$. In $s_j^{\rm succ}(\delta)$, RX$_i$ has VAoI $\tilde v_i$, whereas in
$\sigma(s_i^{\rm succ}(\delta))$, RX$_i$ has VAoI $\tilde v_j$. Since
$v_i<v_j$, we have $\tilde v_i\le \tilde v_j$, and hence
\begin{align}
s_j^{\rm succ}(\delta)
\preceq
\sigma(s_i^{\rm succ}(\delta)).
\end{align}
Using Proposition~\ref{prop:rvf_monotonicity} and then
Proposition~\ref{prop:symmetry},
\begin{align}
U(s_j^{\rm succ}(\delta))
\le
U(\sigma(s_i^{\rm succ}(\delta)))
=
U(s_i^{\rm succ}(\delta)).
\label{eq:serve_oldest_success_value}
\end{align}
For transmission to RX$_i$, we have, 
\begin{align}
\mathbb E
\left[
U(s')
\,\middle|\,
s,i,\delta
\right]
=
pU(s_i^{\rm succ}(\delta))
+
(1-p)U(s^{\rm fail}(\delta)),
\end{align}
and for RX$_j$, we have,
\begin{align}
\mathbb E
\left[
U(s')
\,\middle|\,
s,j,\delta
\right]
=
pU(s_j^{\rm succ}(\delta))
+
(1-p)U(s^{\rm fail}(\delta)).
\end{align}
The failure contribution is identical under both actions, and
\eqref{eq:serve_oldest_success_value} shows that the success contribution under
serving RX$_j$ is no larger than that under serving RX$_i$. Hence,
\begin{align}
\mathbb E
\left[
U(s')
\,\middle|\,
s,j,\delta
\right]
\le
\mathbb E
\left[
U(s')
\,\middle|\,
s,i,\delta
\right].
\label{eq:expecOlder}
\end{align}

\begin{figure*}[t]
    \centering
    \includegraphics[width=0.85\linewidth]{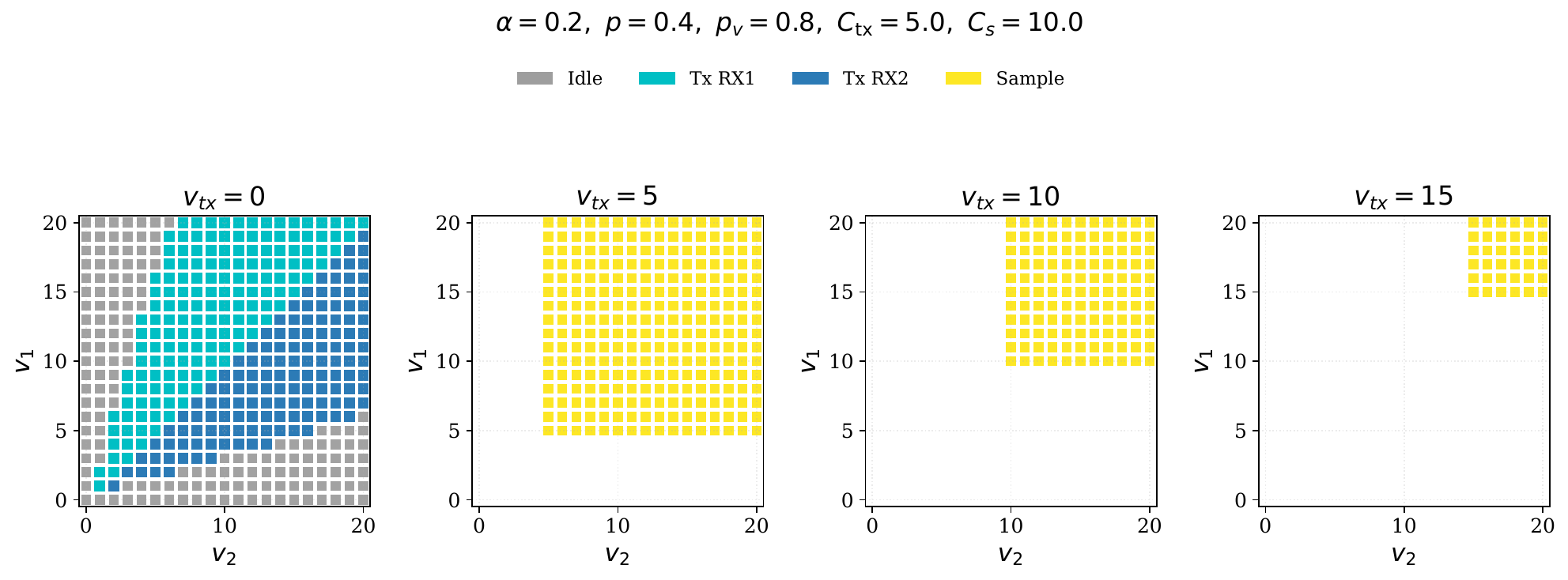}
    \caption{Fixed-$v_{tx}$ slices of the optimal RVI policy over $(v_1,v_2)$ for
the two-receiver case $(n=2)$.On the $v_{tx}=0$ slice for the two-receiver case
$(n=2)$. The policy is symmetric around $v_1=v_2$, and away from the diagonal
the decision switches from serving the larger-VAoI receiver to
idling as the receiver-VAoI difference grows.}
    \label{fig:optimal_policy_slices}
\end{figure*}

Both actions are direct transmissions and therefore incur the same transmission
cost $C_{\mathrm{tx}}$. Hence, by the cost definition in \eqref{eq:cost}, the comparison depends only on the expected VAoI sum.
For a fixed $\delta$, if the transmission to RX$_i$ succeeds, the VAoI sum is,
\begin{align}
\tilde v_{tx}
+
\sum_{r\neq i}\tilde v_r.
\end{align}
If the transmission to RX$_j$ succeeds, the VAoI sum is,
\begin{align}
\tilde v_{tx}
+
\sum_{r\neq j}\tilde v_r.
\end{align}
The difference between them is, 
\begin{align}
\left(
\tilde v_{tx}
+
\sum_{r\neq j}\tilde v_r
\right)
-
\left(
\tilde v_{tx}
+
\sum_{r\neq i}\tilde v_r
\right)
=
\tilde v_i-\tilde v_j
\le 0.
\end{align}
Thus, conditional on success, serving RX$_j$ gives a VAoI sum no
larger than serving RX$_i$. Moreover, when $\delta=0$, we have
$\tilde v_i=v_i<v_j=\tilde v_j$, so the inequality is strict. Since both actions also have the same failure
state, the expected VAoI sum, and hence the cost, is
strictly smaller under serving RX$_j$:
\begin{align}
c(s,j)<c(s,i).
\end{align}
Combining with \eqref{eq:expecOlder}gives,
\begin{align}
Q(s,j)<Q(s,i).
\end{align}
Therefore, among direct transmissions, an optimal action targets a receiver
with maximum VAoI.
\end{proof}

\section{VAoI-Difference Threshold Structure with Minimum VAoI-activation}

The remaining decision is whether to transmit directly to an
oldest receiver or to idle and rely on receiver gossip. By permutation invariance and Proposition~\ref{prop:serve_oldest}, we
assume WLOG that RX$_n$ is the oldest receiver and write
\begin{align}
s_d=(v_{tx},m,\ldots,m,m+d),
\qquad d\ge0.
\end{align}
The relevant comparison is between idling, $u=0$, and directly
transmitting to RX$_n$, $u=n$. Under idling, each of the $n-1$ receivers selects one gossip target
uniformly among the other $n-1$ receivers. Hence, a given receiver
successfully updates RX$_n$ with probability $p_v/(n-1)$. Therefore, the
probability that RX$_n$ receives at least one successful gossip update is
\begin{align}
\tilde p_v
\triangleq
1-\left(1-\frac{p_v}{n-1}\right)^{n-1}.
\end{align}
We define 
\begin{align}
\Delta(d)
\triangleq
Q(s_d,n)-Q(s_d,0).
\end{align}
Thus, direct transmission to RX$_n$ is preferred when $\Delta(d)\le0$, while
idling is preferred when $\Delta(d)>0$.

\begin{proposition}[VAoI-Difference Monotonicity]
\label{prop:age_difference_monotonicity}
Assume symmetric direct and gossip links. If $\tilde p_v>p$, then for fixed
$v_{tx}$ and $m$, $\Delta(d)$ is nondecreasing in $d$.
\end{proposition}

\begin{proof}
Fix $v_{tx}$, $m$, and $d$. For a fixed realization of
$\delta\in\{0,1\}$, let the intermediate VAoIs be given by
\eqref{eq:intermediate_vaoi1}--\eqref{eq:intermediate_vaoi2}. For the state
$s_d=(v_{tx},m,\ldots,m,m+d)$, the first $n-1$ receivers we have, $\tilde m=(m+\delta)\wedge V_{\max}$ while for RX$_n$ we have,  
$\tilde v_n(d)=(m+d+\delta)\wedge V_{\max}$. Thus, $\tilde v_n$ is nondecreasing in $d$. We define,
\begin{align}
x_n(\delta)
&\triangleq
(\tilde v_{tx},\tilde m,\ldots,\tilde m,\tilde v_{tx}),\\
x_g(\delta)
&\triangleq
(\tilde v_{tx},\tilde m,\ldots,\tilde m,\tilde m),\\
x_s(\delta)
&\triangleq
(\tilde v_{tx},\tilde m,\ldots,\tilde m,\tilde v_n).
\end{align}
Here, $x_n(\delta)$ is the state after a successful direct transmission to
RX$_n$, $x_g(\delta)$ is the state where RX$_n$ is refreshed by gossip, and
$x_s(\delta)$ is the state where RX$_n$ is not refreshed.

Under direct transmission to RX$_n$, success occurs with probability $p$.
Therefore, for fixed $\delta$, under direct transmission we have
\begin{align}
\mathbb E
\left[
U(s')
\,\middle|\,
s_d,n,\delta
\right]
=
pU(x_n(\delta))
+
(1-p)U(x_s(\delta)).
\end{align}
Under idling, RX$_n$ receives at least one successful gossip update with
probability $\tilde p_v$. Therefore, we have
\begin{align}
\mathbb E
\left[
U(s')
\,\middle|\,
s_d,0,\delta
\right]
=
\tilde p_v U(x_g(\delta))
+
(1-\tilde p_v)U(x_s(\delta)).
\end{align}
Hence, the difference between them is
\begin{align}
&\mathbb E
\left[
U(s')
\,\middle|\,
s_d,n,\delta
\right]
-
\mathbb E
\left[
U(s')
\,\middle|\,
s_d,0,\delta
\right] \nonumber\\
&=
pU(x_n(\delta))
-
\tilde p_v U(x_g(\delta))
+
(\tilde p_v-p)U(x_s(\delta)).
\label{eq:rv_difference_d}
\end{align}

Next, we compare the expected VAoI sums. Under direct transmission to RX$_n$,
the first $n-1$ receiver VAoIs remain equal to $\tilde m$, while RX$_n$
becomes $\tilde v_{tx}$ with probability $p$ and remains $\tilde v_n$ with
probability $1-p$. Hence,
\begin{align}
\mathbb E
\left[
\sum_{r=1}^n v_r(t+1)
\,\middle|\,
s_d,n,\delta
\right]
=
(n-1)\tilde m
+
p\tilde v_{tx}
+
(1-p)\tilde v_n.
\end{align}
Under idling, RX$_n$ becomes $\tilde m$ with probability $\tilde p_v$ and
remains $\tilde v_n(d)$ with probability $1-\tilde p_v$. Hence,
\begin{align}
&\mathbb E
\left[
\sum_{r=1}^n v_r(t+1)
\,\middle|\,
s_d,0,\delta
\right] \nonumber\\
&=
(n-1)\tilde m
+
\tilde p_v\tilde m
+
(1-\tilde p_v)\tilde v_n(d).
\end{align}
Therefore, the expected VAoI sum difference between direct transmission and
idling is
\begin{align}
&\mathbb E
\left[
\sum_{r=1}^n v_r(t+1)
\,\middle|\,
s_d,n,\delta
\right]
-
\mathbb E
\left[
\sum_{r=1}^n v_r(t+1)
\,\middle|\,
s_d,0,\delta
\right] \nonumber\\
&=
p\tilde v_{tx}
-
\tilde p_v\tilde m
+
(\tilde p_v-p)\tilde v_n.
\label{eq:cost_difference_d}
\end{align}
The direct transmission has additional constant $C_{\mathrm{tx}}$, while the
idling action cost is zero. This does not affect monotonicity in $d$. Now in
\eqref{eq:rv_difference_d} and \eqref{eq:cost_difference_d}, for fixed
$v_{tx}$, $m$, and $\delta$, the terms involving $x_n(\delta)$, $x_g(\delta)$,
$\tilde v_{tx}$, and $\tilde m$ do not depend on $d$. The only $d$-dependent
terms depending on $\tilde v_n$  are
\begin{align}
(\tilde p_v-p)\tilde v_n
\qquad \text{and} \qquad
(\tilde p_v-p)U(x_s(\delta)).
\end{align}
Since $\tilde p_v>p$ and $\tilde v_n$ is nondecreasing in $d$, the first
term is nondecreasing in $d$. Also, $x_s(\delta)$ is componentwise
nondecreasing in $d$, and by Proposition~\ref{prop:rvf_monotonicity},
$U(x_s(\delta))$ is nondecreasing in $d$. Hence, the second term is also
nondecreasing in $d$. Thus, for each realization of $\delta$, the difference
$Q(s_d,n)-Q(s_d,0)$ is nondecreasing in $d$. Taking expectation over $\delta$
preserves monotonicity. Therefore, $\Delta(d)$ is nondecreasing in $d$.

\end{proof}

\begin{figure}[t]
    \centering
    \includegraphics[width=0.8\linewidth]{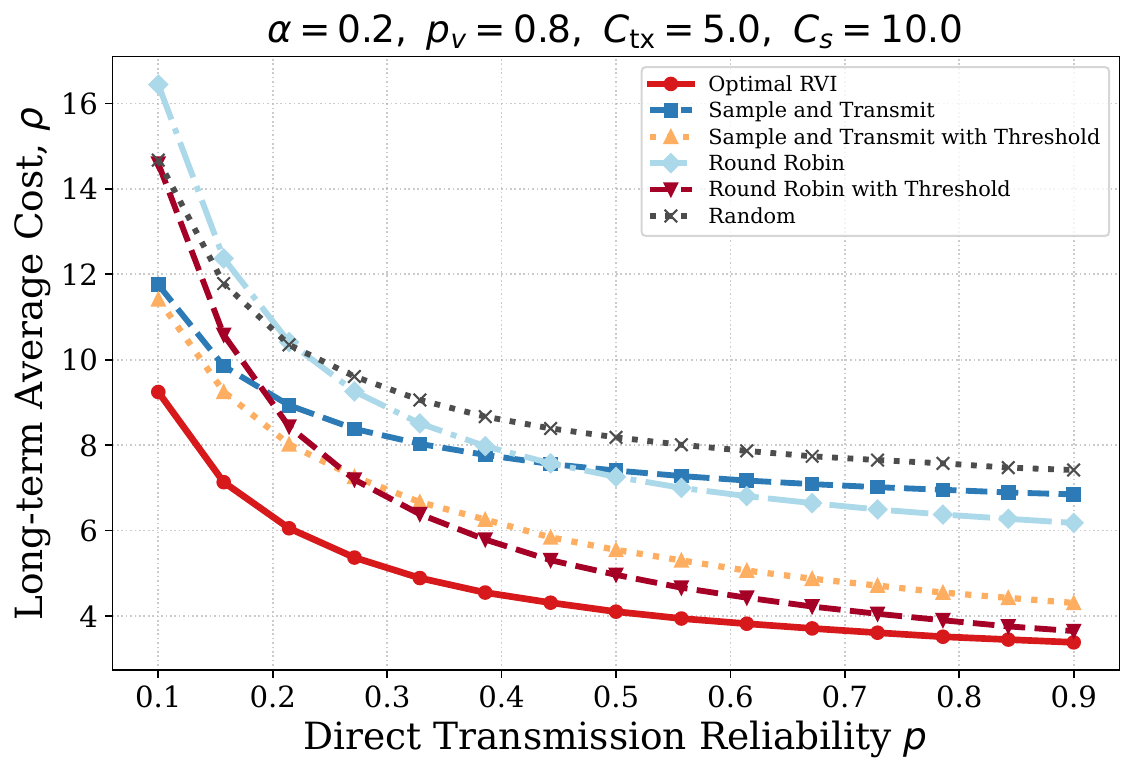}
    \caption{Long-term average cost versus direct-link reliability \(p\).}
    \label{fig:p_sweep}
\end{figure}

\begin{theorem}[VAoI-Difference Threshold]
\label{thm:age_difference_threshold}
Consider the symmetric network and suppose $\tilde p_v>p$. Fix $v_{tx}$ and
$m$. As $d$ increases along 
\begin{align}
s_d=(v_{tx},m,\ldots,m,m+d),\qquad d\ge0,
\end{align}
the decision switches at most once from direct transmission to
RX$_n$ to idling.
\end{theorem}

\begin{proof}
In $s_d$, RX$_n$ has the largest VAoI. Therefore, by
Proposition~\ref{prop:serve_oldest}, if direct transmission is selected, it is
optimal to serve RX$_n$. Hence, the relevant comparison is between direct
transmission to RX$_n$ and idling.

Recall that
\begin{align}
\Delta(d)=Q(s_d,n)-Q(s_d,0),
\end{align}
so direct transmission to RX$_n$ is preferred when $\Delta(d)\le0$, while
idling is preferred when $\Delta(d)>0$. By
Proposition~\ref{prop:age_difference_monotonicity}, $\Delta(d)$ is
nondecreasing in $d$. Therefore, once $\Delta(d)>0$, it remains positive for
all larger $d$. Hence, the decision switches at
most once from direct transmission to RX$_n$ to idling as $d$ increases.
\end{proof}

\begin{corollary}[Minimum-VAoI Activation]
\label{cor:min_age_activation}
Let $m^\star(v_{tx})$ denote the minimum diagonal receiver
VAoI above which direct transmission is preferred to idling. Then
\begin{align}
m^\star(v_{tx})
\le
v_{tx}+\left\lfloor \frac{C_{\mathrm{tx}}}{p}\right\rfloor+1.
\end{align}
\end{corollary}

\begin{proof}
At $s_0=(v_{tx},m,\ldots,m)$, gossip cannot reduce any receiver VAoI because
all receivers have the same VAoI. Hence, idling leads to
$x_m(\delta)=(\tilde v_{tx},\tilde m,\ldots,\tilde m)$. Under direct
transmission to any receiver, say RX$_n$, failure gives the same state
$x_m(\delta)$. On the other hand, successful transmission gives
\begin{align}
x_n(\delta)
=
(\tilde v_{tx},\tilde m,\ldots,\tilde m,\tilde v_{tx}).
\end{align}
For fixed $\delta\in\{0,1\}$, $m+1\le V_{\max}$ and $v_{tx}\le m$ imply that
$\tilde v_{tx}=v_{tx}+\delta$, $\tilde m=m+\delta$, and therefore
$\tilde v_{tx}-\tilde m=v_{tx}-m$. The expected VAoI sum difference between direct transmission and
idling is
\begin{align}
&\left[
(1-p)n\tilde m
+
p\big((n-1)\tilde m+\tilde v_{tx}\big)
\right]
-
n\tilde m \nonumber\\
&\qquad =
p(\tilde v_{tx}-\tilde m).
\end{align}
Therefore, by \eqref{eq:cost},
\begin{align}
c(s_0,n)-c(s_0,0)
=
C_{\mathrm{tx}}
+
p(\tilde v_{tx}-\tilde m).
\label{eq:diag_cost_diff}
\end{align}

To complete the comparison in \eqref{eq:qval}, we now consider the expectation term. Under idling, the
next state is $x_m(\delta)$. Under direct transmission, success gives
$x_n(\delta)$ with probability $p$, and failure gives $x_m(\delta)$ with
probability $1-p$. Hence,
\begin{align}
&\mathbb E[U(s')\mid s_0,n,\delta]
-
\mathbb E[U(s')\mid s_0,0,\delta] \nonumber\\
&\qquad =
p\left[U(x_n(\delta))-U(x_m(\delta))\right].
\label{eq:diag_value_diff}
\end{align}
Since $v_{tx}\le m$, we have $x_n(\delta)\preceq x_m(\delta)$. By
Proposition~\ref{prop:rvf_monotonicity},
$U(x_n(\delta))\le U(x_m(\delta))$, so the difference in
\eqref{eq:diag_value_diff} is nonpositive. Combining \eqref{eq:qval}, \eqref{eq:diag_cost_diff}, and
\eqref{eq:diag_value_diff}, and using
$\tilde v_{tx}-\tilde m=v_{tx}-m$, we obtain
\begin{align}
Q(s_0,n)-Q(s_0,0)
&\le
C_{\mathrm{tx}}
+
p(\tilde v_{tx}-\tilde m) \nonumber\\
&=
C_{\mathrm{tx}}-p(m-v_{tx}).
\end{align}
Thus, direct transmission is strictly preferred to idling whenever
\begin{align}
C_{\mathrm{tx}}-p(m-v_{tx})<0,
\end{align}
or equivalently,
\begin{align}
m-v_{tx}>\frac{C_{\mathrm{tx}}}{p}.
\end{align}
Since VAoI is integer valued, $m-v_{tx}$ is integer valued. Hence, the above
strict inequality is guaranteed when,
\begin{align}
m\ge
v_{tx}+\left\lfloor \frac{C_{\mathrm{tx}}}{p}\right\rfloor+1.
\end{align}
Therefore,
\begin{align}
m^\star(v_{tx})
\le
v_{tx}+\left\lfloor \frac{C_{\mathrm{tx}}}{p}\right\rfloor+1.
\end{align}
\end{proof}

\begin{figure}[t]
    \centering
    \includegraphics[width=0.8\linewidth]{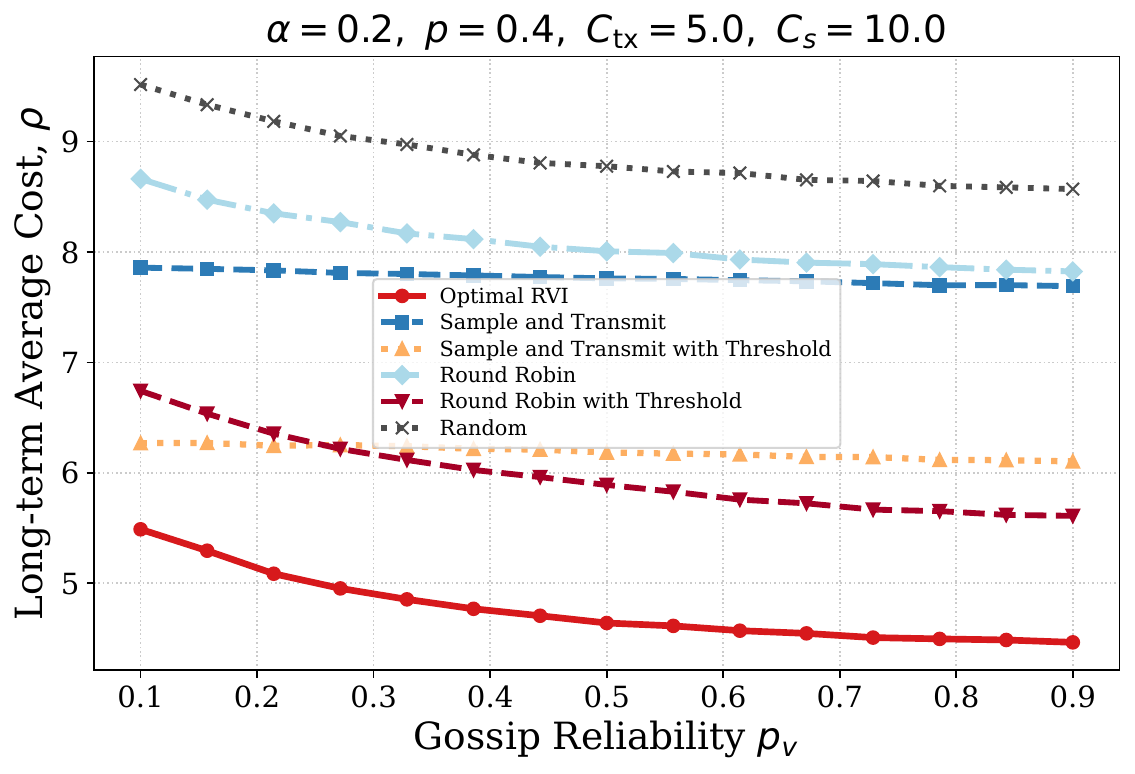}
    \caption{Long-term average cost versus gossip reliability \(p_v\).}
    \label{fig:pv_sweep}
\end{figure}

\section{Numerical Results}

We evaluate the optimal policy using relative value iteration (RVI) with $V_{\max}=20$. We focus on a
two receiver case $n=2$. Although the analytical results are stated for general
$n$, the case $n=2$ allows the full state $(v_{tx},v_1,v_2)$ to be visualized
through 3-D policy plots and fixed-$v_{tx}$ slices. This makes it
possible to directly inspect and get the intuition of structures. For larger
$n$, the same RVI method applies, but the policy lies in an
$(n+1)$-dimensional state space and is not directly visualizable. The symmetric setting is used throughout, so $p_1=p_2=p$ and
$p_{v,12}=p_{v,21}=p_v$. Unless otherwise stated, we use
$
\alpha=0.2,\ p=0.4,\ p_v=0.8,\ C_{\mathrm{tx}}=5,\ C_s=10.
$

We compare the optimal RVI policy with five baselines. The
\emph{Sample and Transmit} policy samples whenever the transmitter VAoI $v_{tx}$ is
positive and otherwise transmits to the older receiver. The
\emph{Sample and Transmit with Threshold} policy uses thresholds for both
sampling and transmission, which naturally incorporates idling. The \emph{Round Robin} policy cycles through
sampling, direct transmission to the older receiver, and idling, while
\emph{Round Robin with Threshold} executes the same cycle subject to sampling
and transmission thresholds, thereby enabling more idling slots. The \emph{Random} policy chooses uniformly among
idling, transmission to RX$_1$, transmission to RX$_2$, and sampling. For the threshold based baselines, the thresholds are selected to provide the
best observed performance, which is observed with thresholds of $1$. 

The policy geometry in Fig.~\ref{fig:optimal_policy_3d} and
Fig.~\ref{fig:optimal_policy_slices} shows three operating regimes. When
$v_{tx}$ is large, sampling dominates because direct transmission would only
propagate a stale transmitter version into the gossip network, matching
Theorem~\ref{thm:sampling_threshold}. When $v_{tx}$ is small, the transmitter
has a useful version, and direct transmission may become attractive as well as idling is another option. The
transmission regions are symmetric around $v_1=v_2$, and the policy serves
RX$_1$ when $v_1>v_2$ and RX$_2$ when $v_2>v_1$, as analyzed in
Proposition~\ref{prop:serve_oldest}. Idling appears where receiver gossip can
improve freshness without paying the transmission cost.

\begin{figure}[t]
    \centering
    \includegraphics[width=0.8\linewidth]{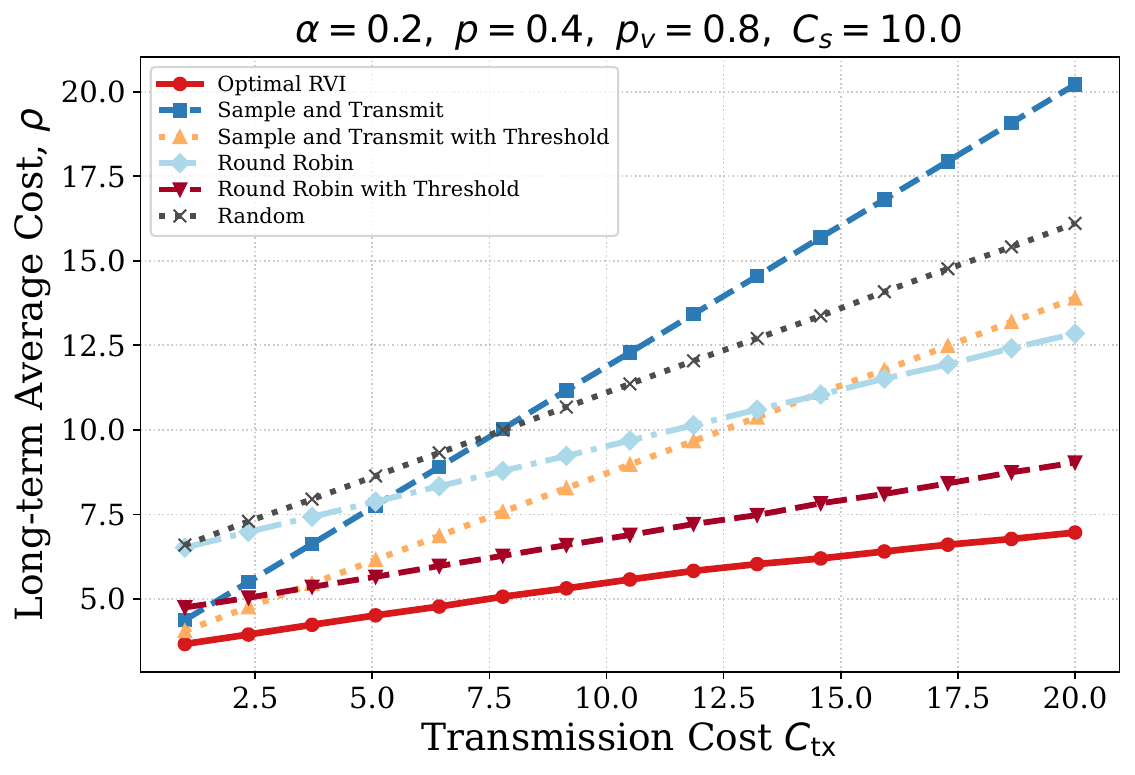}
    \caption{Long-term average cost versus transmission cost
    \(C_{\mathrm{tx}}\).}
    \label{fig:Ctx_sweep}
\end{figure}

\begin{figure}[t]
    \centering
    \includegraphics[width=0.8\linewidth]{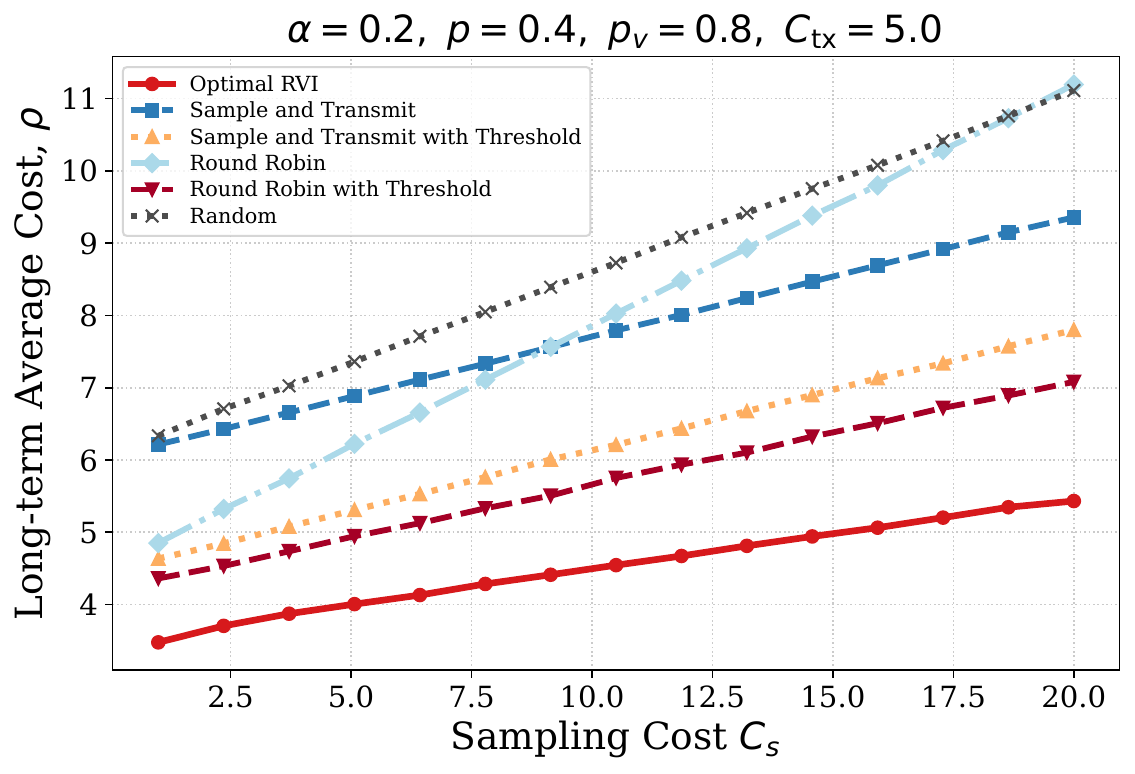}
    \caption{Long-term average cost versus sampling cost \(C_s\).}
    \label{fig:Cs_sweep}
\end{figure}

\begin{figure}[t]
    \centering
    \includegraphics[width=0.8\linewidth]{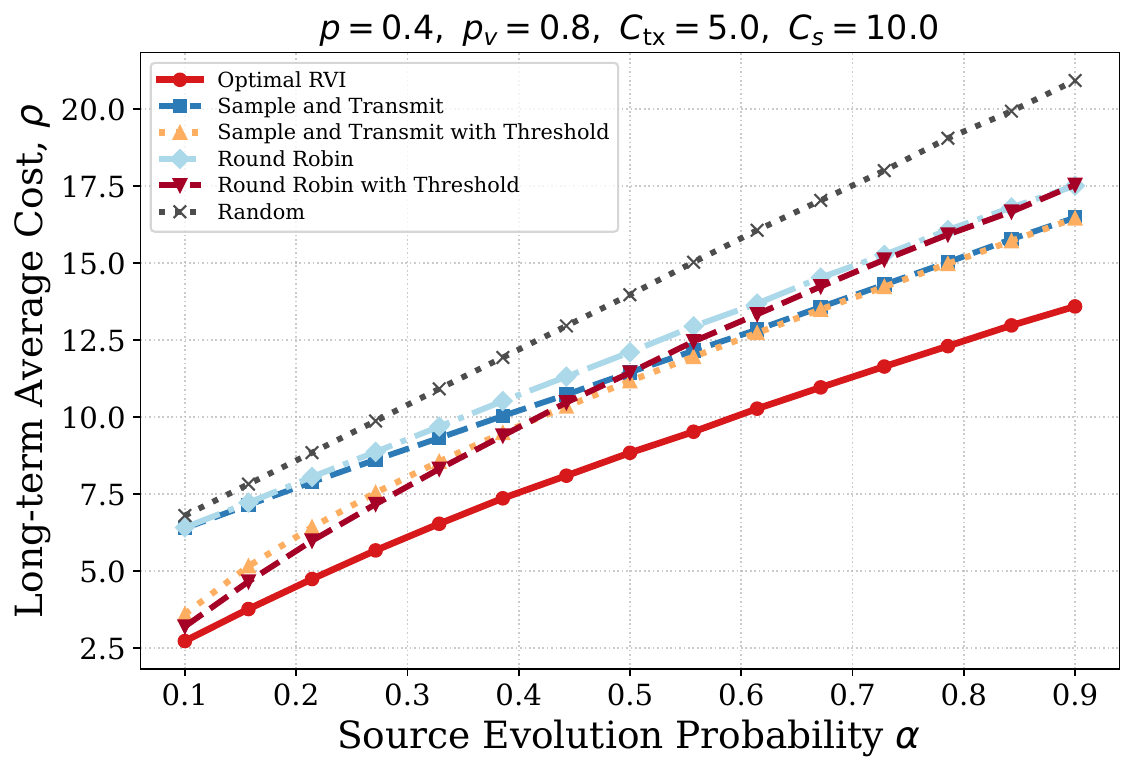}
    \caption{Long-term average cost versus source evolution probability
    \(\alpha\).}
    \label{fig:alpha_sweep}
\end{figure}

We focus on Fig.~\ref{fig:optimal_policy_slices} on the
$v_{tx}=0$ slice. Along the diagonal $v_1=v_2$,  direct transmission
eventually activates as the common receiver VAoI grows, consistent with
Corollary~\ref{cor:min_age_activation}. Away from the diagonal, the policy
first serves the larger VAoI receiver and then switches to idling when the VAoI
difference becomes large enough for gossip to be more valuable, consistent with
Theorem~\ref{thm:age_difference_threshold}. The reliability sweeps in Fig.~\ref{fig:p_sweep} and Fig.~\ref{fig:pv_sweep}
show that direct links and gossip links affect the policy in different ways.
Increasing $p$ makes direct transmission effective and makes the performance of all the policies better. Increasing $p_v$ changes the role of idling. When gossip is
reliable, idling becomes an active freshness action. The optimal policy exploits both effects by adapting when to
sample, transmit, and rely on gossip, which explains its consistent advantage over less adaptive baselines. The cost sweeps in Fig.~\ref{fig:Ctx_sweep} and Fig.~\ref{fig:Cs_sweep}
highlight the economic value of structure. As $C_{\mathrm{tx}}$ increases, baselines degrade because they continue to pay for direct
updates even when gossip or sampling is the smarter cost-effective choice. As $C_s$
increases, the optimal policy delays sampling until transmitter staleness
justifies the cost, which is the operational meaning of the sampling threshold structure. Fig.~\ref{fig:alpha_sweep} shows that larger $\alpha$ increases the long-term
average cost because new source versions are generated more frequently and
stale information accumulates faster. The optimal policy remains lowest by
balancing the actions available in the system. 

Overall, the numerical results show that the optimal policy exploits the
geometry of the VAoI state space rather than following a fixed update pattern.
Sampling is driven by transmitter staleness, direct transmission is activated
when receiver VAoIs are sufficiently large or when one receiver is the best
direct target, and idling becomes valuable when gossip can exploit VAoI
differences between receivers. This explains the consistent performance gain of
the optimal policy across reliability, cost, and source dynamics.

\section{Conclusion}

We studied optimal freshness control in a fully connected gossip network under
the Version Age of Information (VAoI) metric. Unlike conventional models that
assume the transmitter always has the freshest source version, our model
captures transmitter staleness and costly source sampling. This leads to a problem in which the transmitter must decide when to
sample, when to inject its available version through direct transmission, and
when to idle so that receiver gossip can improve freshness at no direct
transmission cost.

The main contribution of this work is a structural characterization of the
optimal policy. We formulated the problem as an infinite-horizon average-cost
MDP and proved monotonicity of the relative value function. This enabled us to
show that sampling has a threshold structure in the transmitter VAoI: for fixed
receiver VAoIs, once sampling becomes optimal, it remains optimal for all
larger feasible transmitter VAoIs. Under symmetric links, the
optimal direct transmission targets a receiver with the largest VAoI. Moreover,
when gossip is sufficiently reliable relative to direct transmission, the
policy can switch from direct transmission to idling as one
receiver becomes much staler than the others. This reveals a key mechanism in
gossip networks that a large VAoI imbalance can make idling valuable because the
fresher receivers can repair the stale receiver through gossip. We further
derived a minimum-VAoI activation condition showing that, on the receiver-VAoI
diagonal, direct transmission eventually dominates idling once the common
receiver VAoI is sufficiently large.

The numerical results show that these structural properties are visible
principles of the optimal policy. The policy visualizations confirm
the sampling threshold, diagonal activation of direct
transmission, and the switch from direct transmission to idling as receiver
VAoI imbalance grows. Parameter sweeps demonstrate that the optimal RVI policy
consistently outperforms natural baselines by adapting sampling, transmission,
and idling decisions to link reliabilities, operation costs, and source
dynamics. Overall, the results show that freshness in gossip networks should be
managed by coordinated use of source sampling, direct injection, and gossip. These structural insights open several directions for future work. The
threshold and monotonicity properties can be used to design low-complexity
heuristics for larger networks where exact dynamic programming is infeasible.
They can guide structure-aware reinforcement learning by reducing the
effective policy search space. Important extensions include heterogeneous
direct and gossip links, multiple sources, time-varying gossip graphs and
energy-constrained transmitters.

\bibliographystyle{IEEEtran}
\bibliography{references_TCOM}
\appendices

\section{Proof of Proposition~\ref{prop:feasibility}}
\label{app:feasibility}

Let $s(t)=(v_{tx},v_1,\ldots,v_n)\in\mathcal S$, where $0\le v_{tx}\le v_i\le V_{\max},\forall i$. It holds that after source evolution in the intermediate state, $0\le \tilde v_{tx}\le \tilde v_i\le V_{\max},\forall i$. If $u=i$, then upon success RX$_i$ is set to $\tilde v_{tx}$, while every $j\neq i$ remains at $\tilde v_j$. Upon failure,
all receivers remain at $\tilde v_i, \forall i$. In both cases, every receiver VAoI is between $\tilde v_{tx}$ and $V_{\max}$. If $u=0$, the transmitter idles and $v_{tx}(t+1)=\tilde v_{tx}$. Receiver $i$
updates as \eqref{eq:gossip_individual}.
Since $\tilde v_i\ge \tilde v_{tx}$, $\forall i$. Hence, $v_i(t+1)\ge v_{tx}(t+1),\forall i$. So, every receiver VAoI is between $\tilde v_{tx}$ and $V_{\max}$. If $u=n+1$, sampling resets the transmitter VAoI to zero, and the receivers
gossip as in \eqref{eq:gossip_individual} which gives
$0=v_{tx}(t+1)\le v_i(t+1) \le V_{max},\forall i$. Thus, for every action and transition outcome,
$0\le v_{tx}(t+1)\le v_i(t+1)\le V_{\max},
\forall i$.
and therefore $s(t+1)\in\mathcal S$. Hence, $\mathcal S$ is closed.
\hfill $\blacksquare$

\section{Proof of Proposition~\ref{prop:weak_comm}}
\label{app:weak_comm}

Let $s=(v_{tx},v_1,\ldots,v_n)$ and
$s'=(v_{tx}',w_1,\ldots,w_n)$ be arbitrary feasible states. Consider a
stationary randomized policy with $\pi(u|s)>0$ for all
$u\in\mathcal A$ and $s\in\mathcal S$. First, we show that $s^0=(0,0,\ldots,0)$ is accessible from $s$. Choose the
sampling action $u=n+1$, and realize $\delta=0$ with no successful receiver
gossip. This event has positive probability since $1-\alpha>0$ and
$1-p_{v,ji}>0$ for all $j\neq i$. It resets the transmitter VAoI to zero while
leaving the receiver VAoIs unchanged. Then, over the next $n$ slots, transmit
once to each receiver RX$_i$, realizing $\delta=0$ and a successful direct
transmission in each slot. These outcomes have probability
$(1-\alpha)^n\prod_{i=1}^n p_i>0$. Since the transmitter VAoI remains zero,
each receiver is reset to zero. Hence, $s^0$ is reached with positive
probability. Next, we show that $s'$ is accessible from $s^0$. Define
\begin{align}
d_i=w_i-v_{tx}',\qquad i=1,\ldots,n.
\end{align}
Choose an ordering $\ell_1,\ldots,\ell_n$ of the receivers such that
$d_{\ell_1}\ge d_{\ell_2}\ge\cdots\ge d_{\ell_n}$. Starting from $s^0$, reset
RX$_{\ell_1}$ to VAoI zero by choosing a direct transmission to RX$_{\ell_1}$,
realizing $\delta=0$, and realizing a successful transmission. This event has
probability $(1-\alpha)p_{\ell_1}>0$. Then create
$d_{\ell_1}-d_{\ell_2}$ slots in which $\delta=1$ occurs and the chosen direct
transmission fails. Each such slot has probability $\alpha(1-p_\ell)>0$ for
some chosen receiver $\ell$, and increases every VAoI by one without refreshing any receiver. Next, sample with $\delta=0$ and no successful gossip, so that $v_{tx}=0$, and
reset RX$_{\ell_2}$ to VAoI zero by a successful direct transmission with
$\delta=0$. Continuing similarly, after resetting RX$_{\ell_j}$ to VAoI zero,
create $d_{\ell_j}-d_{\ell_{j+1}}$ slots in which $\delta=1$ occurs and the
chosen direct transmission fails, for $j=1,\ldots,n-1$. Each receiver reset has
probability $(1-\alpha)p_{\ell_j}>0$, and each such VAoI-increment slot has
probability $\alpha(1-p_\ell)>0$. After resetting RX$_{\ell_n}$, create
$d_{\ell_n}$ additional such slots. Thus, RX$_{\ell_j}$ experiences
$
(d_{\ell_j}-d_{\ell_{j+1}})+\cdots+
(d_{\ell_{n-1}}-d_{\ell_n})+d_{\ell_n}=d_{\ell_j}
$
VAoI increments after its reset. Hence every receiver RX$_i$ ends with VAoI
$d_i$. At this point the transmitter VAoI may be positive. Choose the sampling action
once more with $\delta=0$ and no successful gossip. This event has positive
probability and resets only the transmitter VAoI to zero while leaving the
receiver VAoIs unchanged. Hence, $(0,d_1,\ldots,d_n)$ is reached with positive probability. Finally, from $(0,d_1,\ldots,d_n)$, create $v_{tx}'$ slots in which
$\delta=1$ occurs and the chosen direct transmission fails. These slots have
positive probability and increase all VAoIs by $v_{tx}'$ without refreshing any
receiver, yielding
\begin{align}
(v_{tx}',d_1+v_{tx}',\ldots,d_n+v_{tx}')
=
(v_{tx}',w_1,\ldots,w_n)
=
s',
\end{align}
since $d_i=w_i-v_{tx}'$. Therefore, $s'$ is accessible from $s^0$. Thus every feasible state can reach $s^0$, and $s^0$ can reach every feasible
state. Hence all states in $\mathcal S$ communicate under the randomized policy,
and the MDP is communicating.
\hfill $\blacksquare$
\section{Proof of Lemma~\ref{lem:order_preservation}}
\label{app:order_preservation}

Let $s\preceq s'$, where
$
s=(v_{tx},v_1,\ldots,v_n)
$
and
$
s'=(v'_{tx},v'_1,\ldots,v'_n).
$
For any fixed $\delta\in\{0,1\}$, by \eqref{eq:intermediate_vaoi1} and \eqref{eq:intermediate_vaoi2} the intermediate states satisfy
$\tilde s\preceq \tilde s'$. Consider a direct transmission $u=i$, $i\in\{1,\ldots,n\}$.
If the transmission fails, the next states are $\tilde s$ and $\tilde s'$,
so the order is preserved. If the transmission succeeds, then RX$_i$ is set
to the transmitter VAoI in both systems. Hence, the $i$-th receiver coordinates
in the two next states are $\tilde v_{tx}$ and $\tilde v'_{tx}$, respectively.
Since $\tilde v_{tx}\le \tilde v'_{tx}$, and all other coordinates are unchanged from the intermediate states, therefore $\tilde s\preceq \tilde s'$. Thus,
$
F_{i,\omega,\delta}(s)\preceq F_{i,\omega,\delta}(s')
$. Next we consider the idle action $u=0$. The transmitter coordinates remain
$\tilde v_{tx}$ and $\tilde v'_{tx}$. For each
receiver, under the fixed gossip outcome $\omega$, by
\eqref{eq:gossip_individual} and the monotonicity of the minimum operation, $\tilde v_{i} \le \tilde v'_{i}$ holds for $i \in \{1,2,\ldots, n\}$. Hence,
$
F_{0,\omega,\delta}(s)\preceq F_{0,\omega,\delta}(s')
$. Finally, consider the sampling action $u=n+1$. The transmitter VAoI is reset
to zero in both systems, so $v_{tx} = 0 $ and $v'_{tx} = 0$. The receiver coordinates follow the same gossip update
rule as in \eqref{eq:gossip_individual}. Therefore, by the same
argument,$\tilde v_{i} \le \tilde v'_{i}$ holds for $i \in \{1,2,\ldots, n\}$.
$
F_{n+1,\omega,\delta}(s)\preceq F_{n+1,\omega,\delta}(s')
$. Thus, for every fixed action $u$, source increment $\delta$, and transition
outcome $\omega$, we have:
\begin{align}
    F_{u,\omega,\delta}(s)\preceq F_{u,\omega,\delta}(s')
\end{align}.
\hfill $\blacksquare$
\section{Proof of Lemma~\ref{lem:cost_monotonicity}}
\label{app:cost_monotonicity}

Since the action cost $C(u)$ in the immediate cost \eqref{eq:cost} is
independent of the state, it is enough to compare the expected next receiver
VAoI sums. Let $s\preceq s'$ and fix an action $u\in\mathcal A$. For every fixed source
increment $\delta$ and transition outcome $\omega$, Lemma~\ref{lem:order_preservation}
gives
\begin{align}
F_{u,\omega,\delta}(s)
\preceq
F_{u,\omega,\delta}(s').
\end{align}
Therefore, under the same realization $(u,\delta,\omega)$, each next receiver
VAoI starting from $s$ is no larger than the corresponding next receiver VAoI
starting from $s'$. Hence,
\begin{align}
\sum_{i=1}^n v_i(t+1)
\le
\sum_{i=1}^n v'_i(t+1),
\end{align}
where $v_i(t+1)$ and $v'_i(t+1)$ denote the next receiver VAoIs generated from
$s$ and $s'$, respectively, under the same realization. Since this inequality holds for every realization of $\delta$ and $\omega$,
taking expectation gives
\begin{align}
\mathbb E
\left[
\sum_{i=1}^n v_i(t+1)
\,\middle|\,
s,u
\right]
\le
\mathbb E
\left[
\sum_{i=1}^n v_i(t+1)
\,\middle|\,
s',u
\right].
\end{align}
Adding the same action cost $C(u)$ to both sides and using \eqref{eq:cost}
gives
\begin{align}
c(s,u)\le c(s',u).
\end{align}

\hfill $\blacksquare$
\end{document}